\documentclass{aa}

\usepackage{txfonts}
\usepackage{graphicx}
\usepackage{natbib}
\bibpunct{(}{)}{;}{a}{}{,}

\def\fhleo{HD\,96273}
\def\bd{BD+07\,2411B}
\def\kms{km\,s$^{-1}$}
\def\mag{$^m$}

\begin{document}

\title{Outbursts on normal stars\thanks{Based on observations collected at the La Silla Observatory, ESO (Chile), with the FEROS
                                         spectrograph at the 2.20 MPI/ESO telescope.}}
\subtitle{FH\,Leo misclassified as a novalike variable}

\author{T. H. Dall\inst{1}
        \and
        L. Schmidtobreick\inst{1}
	\and
	N. C. Santos\inst{2}
	\and
	G. Israelian\inst{3}
}
\offprints{T. H. Dall, \email{tdall@eso.org}}
\institute{
        European Southern Observatory, Casilla 19001, Santiago 19, Chile
	\and
	Centro de Astronomia e Astrof\'isica da Universidade de Lisboa, Observat\'orio Astron\'omico de Lisboa,
	Tapada da Ajuda, 1349-018 Lisboa, Portugal
	\and
	Instituto de Astrof\'isica de Canarias, 38205 La Laguna, Tenerife, Spain
}
\date{Received / Accepted }

\abstract{
We present high resolution spectroscopy of the common proper
motion system FH\,Leo (components \fhleo\ and \bd), which has been classified as a novalike  variable due to an 
outburst observed by Hipparcos, and we present and review the available photometry. We show from our spectra that neither star can possibly be a 
cataclysmic variable, instead they are perfectly normal late-F and 
early-G stars. We measured their radial velocities and derived the atmospheric fundamental parameters,
abundances of several elements including Fe, Ni, Cr, Co, V, Sc, Ti, Ca and Mg, and we derive the age of the system. 
From our analysis we conclude that the stars do indeed constitute a 
physical binary.
However, 
the observed outburst cannot be readily explained.
We examine several explanations, including pollution with scattered light from Jupiter,  binarity, microlensing, background supernovae,
interaction with unseen companions and planetary engulfment.
While no explanation is fully satisfactory, the scattered light and star-planet interaction scenarios emerge as the least unlikely ones, and
we give suggestions for further study.

}

\maketitle

\section{Introduction}
FH\,Leo is a wide visual binary consisting of \object{HD 96273} and \object{BD+07 2411B}. 
The components form a common proper motion (CPM) star pair
separated by 8.31$^{\prime\prime}$
with 98\% probability of being a physical binary according to
\citet{halbwachs1986}. The system has been observed with Hipparcos and has been classified 
as variable and as a probably novalike system (designation NL:) by \citet{kazarovets+1999},
based on the outburst appearance of its Hipparcos lightcurve.

This would have been the first cataclysmic variable (CV) found to be part
of a multiple star system and its study would have cast light on formation and
evolution processes of binary stars.  For these reasons we decided to obtain spectra of both components
of the system in order to start an investigation into the nature and variability of these objects. 

According to SIMBAD it is the brighter
northern source, \fhleo, which\ is the variable star, while \citet{downes+2001} claim 
that it is not yet clear which of the two stars is actually variable.
Also the Hipparcos archive \citep{esa1997} lists both \fhleo\ and \bd\ under the name \object{HIP 54268}, and indeed both stars were included in
the 38\arcsec\ aperture of the Hipparcos detectors. Hence, it is not clear from the Hipparcos data either on which star the outburst took place.
In Sect.~\ref{hipparcos} we present and review the photometry collected by Hipparcos and by the American Association of Variable Star Observers
\citep[AAVSO;][]{mattei2004}.

Our observations are presented in Sect.~\ref{observations}. The analysis of the two stars is given in Sect.~\ref{analysis}, where we
show from our spectroscopy that neither \fhleo\ nor \bd\ can possibly be 
CVs; rather they are both perfectly normal late-F and early-G stars.
In Sect.~\ref{outburst} we discuss the possible
mechanisms that may have caused an outburst on a normal main sequence star.
Finally, in Sect.~\ref{summary} we provide a summary and some concluding remarks concerning further study.

\section{The variability of \fhleo\ and \bd}\label{hipparcos}
We have examined the longterm lightcurves, both of the Hipparcos-archive \citep{esa1997} and
of the AAVSO \citep{mattei2004}. These reveal a constant brightness from JD2447880 to JD2452437,
except for a 0.3$^m$ bright outburst between 
JD2448624 and JD244868.
This outburst was observed by Hipparcos and is the actual reason for
classifying the system as a novalike variable \citep{kazarovets+1999}. The AAVSO lightcurve covers the period JD2451966 to JD2452437, and does not
show any signs of outbursts or any other type of variability.

\begin{figure*}
\centering
\includegraphics[width=17cm]{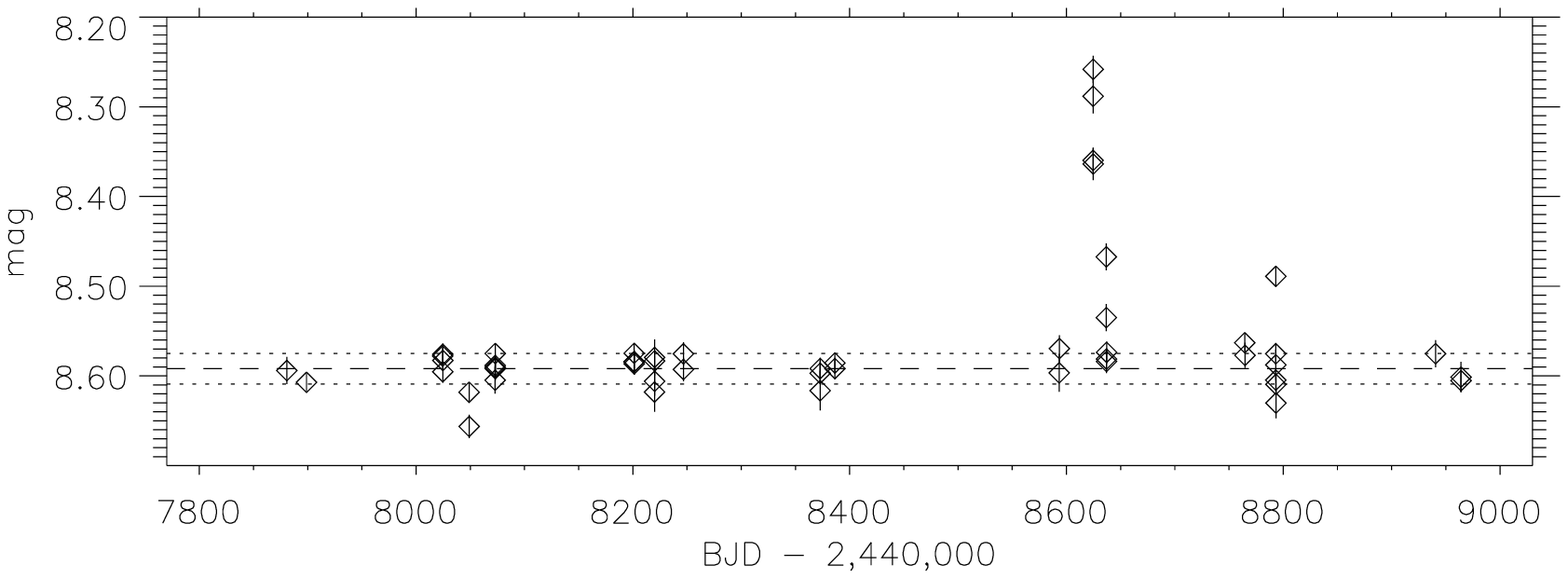}
\includegraphics[width=17cm]{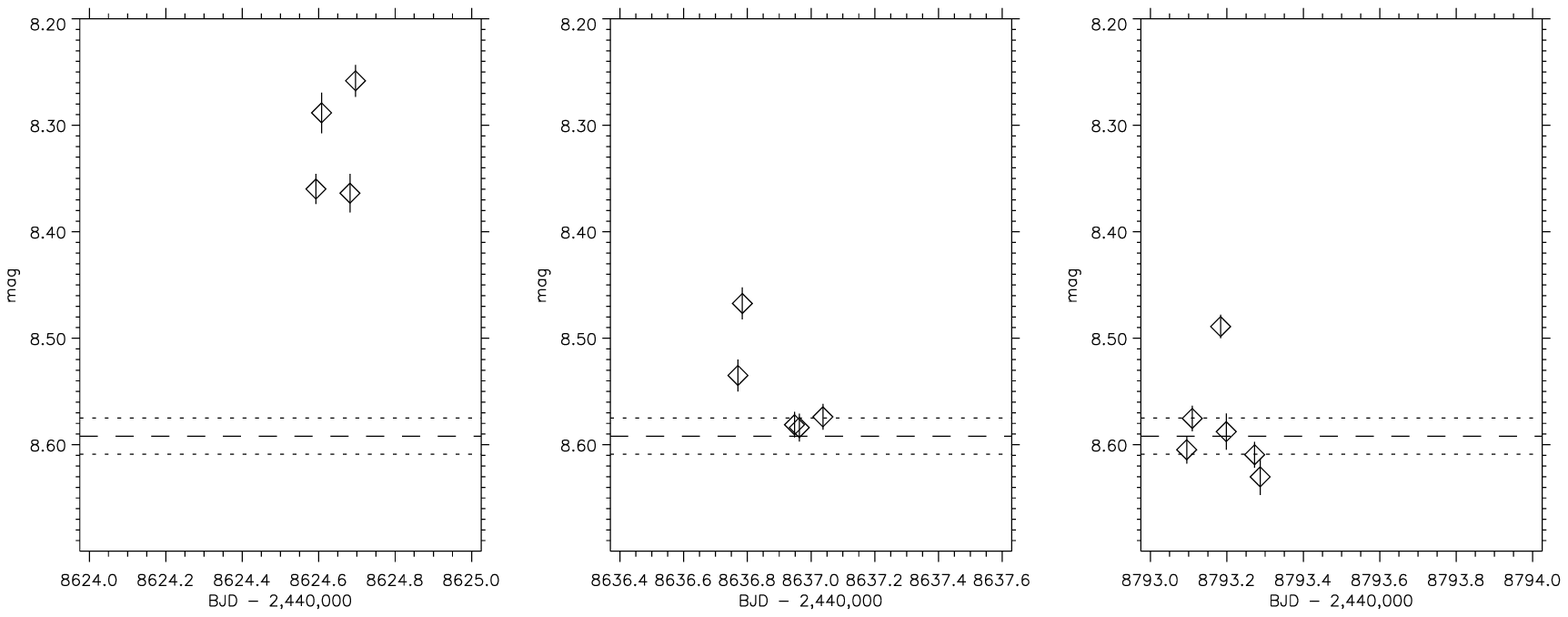}
\caption{Full Hipparcos lightcurve for FH Leo. The dashed line is the mean instrumental magnitude
(not including the outburst), while the dotted lines are the $1\sigma$ standard deviations around the mean. The error bars are Hipparcos
intrinsic errors. The lower plots show a zoom-in on the main event (left and central plot) and one minor event (right plot).
Each of the three lower plots cover one day.}
\label{fig:hip}
\end{figure*}

In Fig.~\ref{fig:hip} we show the Hipparcos data.
 The rise-time might have been very fast
while the decay probably lasted at least 13 days with a possible second event about 170~days later.
The data seem to have a very large spread at the two covered intervals on the decay slope (lower left and central plot of Fig.~\ref{fig:hip}).
This could indicate a non-constant energy production rate during the outburst, which would require some inhomogeneous region or process.
If the system had indeed contained a CV, this short term variation might have been interpreted as 
flickering in the accretion disk. No obvious explanation exists for the phenomenon occurring
in a normal stellar atmosphere, although this does point to
very localized phenomena, e.g. clumps of material ingested in the atmosphere causing hot spots or repeated flare-like outbursts.

\section{Observations}\label{observations}

We have obtained high-resolution (R$\sim$48000) spectra of \fhleo\ and \bd\ with the FEROS spectrograph
at the ESO/MPI-2.20m telescope at ESO's La Silla Observatory, Chile on the night of 2004-01-10.
Standard data reduction was performed with the FEROS DRS pipeline, which is a MIDAS tool
performing bias and flatfield correction, determination of the
wavelength solution, order extraction and wavelength calibration.
The spectra cover the range 3800~--~9000~\AA .
No spectrophotometric standard was observed during this night.
\begin{figure}
\resizebox{\hsize}{!}{\includegraphics{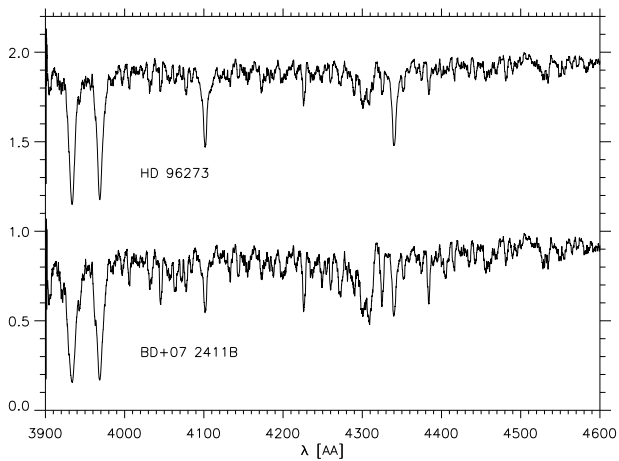}}\\
\caption{\label{spectra} Part of the spectra of \fhleo\
and \bd,  both re-binned to R $\sim$ 400 to simulate
classification resolution. The late--F / early--G type classification
is obvious. Arbitrary flux units.}
\end{figure}

In Fig.~\ref{spectra} we show the spectra of the two stars, re-binned in order
to properly display the usual classification range of 3900~--~4600~\AA, revealing two
seemingly normal stars.

\section{Analysis}\label{analysis}
\begin{figure}
\resizebox{\hsize}{!}{\includegraphics{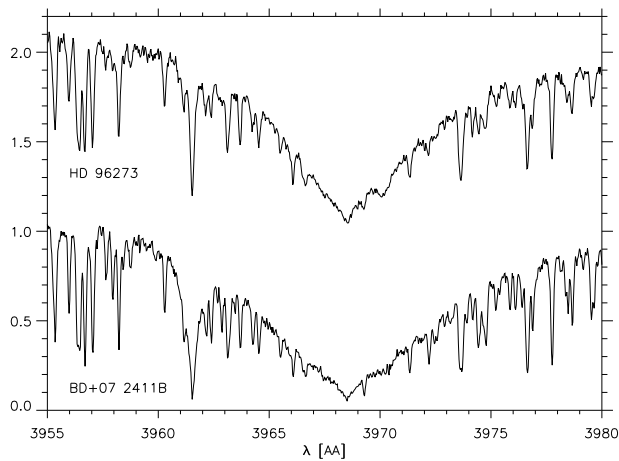}}
\caption{The \ion{Ca}{ii} H line in \fhleo\ and \bd . As can be seen no emission cores are visible, ruling out strong magnetic activity.}
\label{fig:caII}
\end{figure}

The spectra of the two stars resemble normal late-F to early-G type stars with
average rotation rates.
There are no emission lines, especially the \ion{Ca}{ii} H and K lines are devoid of emission cores (see Fig.~\ref{fig:caII}), ruling out
the possibility of strong magnetic activity. There are also no indications that either of the stars should be a 
spectroscopic binary.
We can definitively rule out that any of these two stars
is a cataclysmic variable, and it is indeed doubtful if they are variable at all.

The spectrum of \fhleo\ confirms
the old F8 classification in the Henry Draper Catalog, while no
classification has yet been published for the
fainter companion.  Only a B magnitude of 10.6 (SIMBAD) and a photographic magnitude
of 10.4 \citep{halbwachs1986} are listed for \bd.

\subsection{Atmospheric parameters and abundances}\label{abundances}
We have conducted  a detailed spectral  analysis on both stars, determining the fundamental atmospheric
parameters through the following steps:  Measurement of the equivalent widths (EWs) of the lines over the full spectral region,
determination of a first guess at the basic parameters of the star, calculation of the appropriate model
atmosphere,  and finally the abundance analysis. The method is the same as employed by \citet[][in preparation]{dall+2005}, which
is adopted from the procedures of \citet{morel+2003,bruntt+2002} and \citet{bruntt+2004}.

The first step is the measurement of the EWs, which is accomplished using DAOSPEC\footnote{DAOSPEC has been written
by P.~B.~Stetson for the Dominion Astrophysical Observatory of the Herzberg Institute of Astrophysics, National 
Research Council, Canada.}
\citep[][in preparation]{stetson+pancino2004}, which uses an iterative Gaussian fitting and subtraction procedure to fit the lines
and the effective continuum. The lines are identified using a list of lines from the VALD database 
\citep[][]{kupka+1999,piskunov+1995}, where all lines deeper than 1\% of the continuum are included.  Different 
line-lists for different spectral types can be retrieved directly from the database.
 For our FEROS spectra we used only lines in the region 5000--6800\,\AA, since
for bluer wavelengths the continuum determination becomes uncertain, while beyond 6800\,\AA\ the merging of
orders and hence the continuum shape proved problematic.
 Next, an initial estimate of T$_{\mathrm{eff}}$ is found using the line depth
ratios calibrated by \citet{kovtyukh+2003}.  
 A more accurate 
determination of T$_{\mathrm{eff}}$ will be derived later in the process.
 We adopt $\log g$ and microturbulence parameter $\xi_t$ 
for a canonical ZAMS star, and assume solar
metallicity as our starting point. With this we then calculate the initial model, using the ATLAS9 code
adapted for Linux \citep{kurucz1993,sbordone+2004}.  With the measured EWs and the model, the abundances of the Fe-lines are
calculated using the WIDTH9 code \citep[][modified for PC by V.~Tsymbal]{kurucz1993}, and compared line-by-line
to Solar abundances, calculated from a high S/N solar spectrum obtained 
with the same instrumental setup, and reduced in the same way as the spectra of \fhleo\ and \bd.  This last step is crucial to avoid problems
due to uncertain $gf$-values.
The parameters of the model used to calculate the solar abundances are $T_{\mathrm{eff}}=5778$~K, $\log g = 4.44$,
and $\xi_t = 1.2$~km\,s$^{-1}$.
  Finally, a sigma clipping can be applied to eliminate abundance values deviating e.g.\ because
of errors in the EW determination or due to wrong line identifications.

For the initial $T_\mathrm{eff}$ estimate for \bd\ we used 19 line ratios, which yielded $T_\mathrm{eff} = 5952 \pm 107$~K.
For \fhleo\ the line ratios gave a $T_\mathrm{eff}$ just outside of the valid range of the calibration, hence the initial
temperature guess  was assumed to be just outside the range at 6150~K.

Now the model parameters ($T_{\mathrm{eff}}$, $\log g$,  $\xi_t$, [Fe/H], and [$\alpha$/Fe]) are iteratively modified 
until consistency is reached, defined by the following criteria:  (1) That there are no trends of \ion{Fe}{i} abundance with
EW, wavelength or excitation potential, (2) that the abundances derived from \ion{Fe}{i} and \ion{Fe}{ii} are
the same, (3) that the derived metallicity and $\alpha$-element abundances are consistent with the input model.

 The final model parameters adopted for the two stars
are summarized in Table~\ref{tab:params},  in the rows marked \emph{(a)}, along with the measured $v \sin i$ and radial velocities (RVs).
\begin{table*}
\centering
\caption{\label{tab:params}The derived parameters for \fhleo\ and \bd\ used for the abundance analysis. Note that the
values of $v \sin i$ are about equal to the resolution of the spectrograph, hence they must be regarded as upper
limits. Letters \emph{(a)} and \emph{(b)} corresponds to the two independent analysis discussed in the text.}
\begin{tabular}{lcrrrrrr} \hline
Star  & & $T_\mathrm{eff}$ [K] & $\log g$ & $\xi_t$ [\kms] & $v \sin i$ [\kms] &  RV [\kms] & [Fe/H]  \\ \hline
\fhleo & {\small\emph{(a)}} & 6450$\pm$70 & 4.26$\pm$0.19 & 2.4$\pm$0.2   & 6.7$\pm$0.5 & $-$3.95$\pm$0.44 & $-$0.25$\pm$0.09 \\  
       & {\small\emph{(b)}} & 6493$\pm$99 & 4.50$\pm$0.19 & 2.38$\pm$0.74 &             &                  & $-$0.21$\pm$0.10 \\[0.2ex]
\bd\  & {\small\emph{(a)}} & 5875$\pm$55 & 4.52$\pm$0.11 & 1.4$\pm$0.1    & 5.2$\pm$0.5  & $-$3.19$\pm$0.37 & $-$0.21$\pm$0.07 \\ 
      & {\small\emph{(b)}} & 5849$\pm$54 & 4.45$\pm$0.13 & 1.23$\pm$0.13  &              &                  & $-$0.22$\pm$0.07 \\
 \hline
\end{tabular}
\end{table*}
For the determination of the atmospheric parameters of \bd\ we used 327 \ion{Fe}{i} and 27 \ion{Fe}{ii} lines, 
while for \fhleo\  224 \ion{Fe}{i} and 24 \ion{Fe}{ii} lines
were in common with the solar spectrum.  The RVs were derived using DAOSPEC, while $v \sin i$ was measured
using synthetic spectra calculated with the SYNTH code \citep{piskunov1992,valenti+1996} as implemented  in 
the VWA software \citep{bruntt+2002,bruntt+2004}.

To check our results we conducted an independent analysis following the
procedure outlined in \citet{gonzalez+1996} and \citet{santos+2004},
using the line-list employed by Santos et al., a grid of Kurucz 
ATLAS9 models, and the 2002 version of the radiative transfer 
code MOOG \citep{sneden1973}. The results show that the two methods agree 
within the errors, with values of $T_{\mathrm{eff}} = 6493 \pm 99$~K,
$\log{g} = 4.50 \pm 0.19$, 
$\xi_t = 2.38 \pm 0.74$~km\,s$^{-1}$ and [Fe/H]~=~$-0.21 \pm 0.10$ for
\fhleo\ and $T_{\mathrm{eff}} = 5849 \pm 54$~K,
$\log{g} = 4.45 \pm 0.19$, 
$\xi_t = 1.23 \pm 0.13$~km\,s$^{-1}$ and [Fe/H]~=~$-0.22 \pm 0.07$ for
\bd\ (rows marked \emph{(b)} in Table~\ref{tab:params}).   

The result of both methods agree very well, showing that within the errors the [Fe/H] are the same for the two stars, although
there may be hints at a generally slightly higher metallicity in \bd .

\begin{figure}
\resizebox{\hsize}{!}{\includegraphics{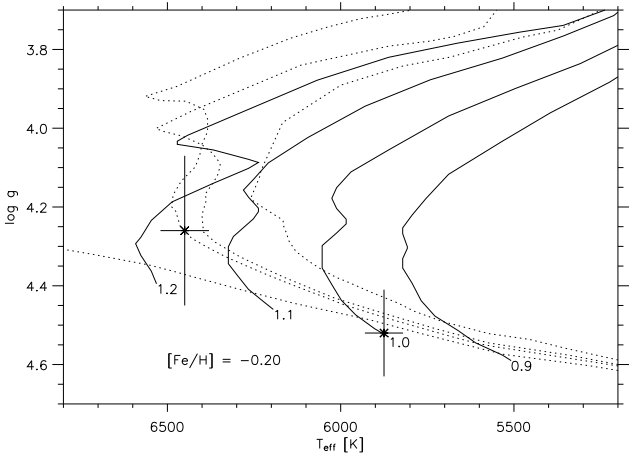}}
\caption{Evolutionary tracks (solid lines) for metallicity
[Fe/H]~=~$-0.20$, 
based on \citet{girardi+2000}. Also shown are the
isochrones (dotted lines) for 1, 2, 3 and 5 Gyr. 
The positions of \fhleo\ and \bd\ are indicated.}
\label{fig:tracks}
\end{figure}

In Fig.~\ref{fig:tracks} we show the positions of the two stars among evolutionary tracks and isochrones calculated
by interpolating in the grid provided by \citet{girardi+2000}.
From this we estimate an age of $\sim 2$~Gyr for \fhleo; an age which within the errors is consistent for \bd\ as well.
However, a word of caution must be given regarding the uncertainties in the temperature and metallicity evolution of the 
models used to compute the isochrones. As demonstrated by \citet{pont+2004} the isochrones of \citet{girardi+2000} tend to overestimate 
the age of the star. Nevertheless, adopting an age estimate of 2~Gyr does not seem unreasonable, and is largely compatible with the lithium age
estimate (Sect.~\ref{li-abund}), especially taking into account the error bars.

\label{abund}
A detailed abundance analysis of the two stars were then done, using the derived stellar parameters
and the corresponding models.  In order to minimize systematic errors due to erroneous $\log gf$ values, we
have calculated all abundances relative to the Sun on a line-to-line basis where corresponding lines are found, and relative
to the values given by \citet{grevesse+1998} where the line is not found in the solar spectrum.

\begin{table}
\centering
\caption{\label{tab:abund}The derived abundances ([M/H]) for \fhleo\ and \bd\ relative to the measured abundances in a Solar spectrum, or relative to
\citet{grevesse+1998} (marked by *). The second column for each star lists [M/Fe], i.e. the abundance within the star of each element relative to iron.}
\begin{tabular}{llr|lr} \hline
Element & \multicolumn{2}{c}{\fhleo} & \multicolumn{2}{c}{\bd} \\ 
      & [M/H] & [M/Fe] & [M/H] & [M/Fe] \\ \hline
Li &     &      & $+0.87(10)$ * & 1.08 \\ 
Mg & $-$0.25(15) & 0.00  & $-$0.10(15) & 0.11 \\
Ca & $-$0.17(11) & 0.08 & $-$0.15(08)  & 0.06 \\
Sc & $-$0.23(10) & 0.02 &  $-$0.13(08) & 0.08 \\
Ti & $-$0.19(12) & 0.06 & $-$0.16(09)  & 0.05 \\
V  & $-$0.19(14) & 0.06 & $-$0.19(11)  & 0.02 \\
Cr & $-$0.25(11) & 0.00 & $-$0.22(10)  & -0.01 \\
Fe & $-$0.25(10) & & $-$0.21(09) & \\
Co & $-$0.24(19) & 0.01 & $-$0.20(11) & 0.01 \\
Ni & $-$0.28(13) & -0.03 & $-$0.24(08) & -0.03 \\
\hline
\end{tabular}
\end{table}
The results of the abundance analysis are given in Table~\ref{tab:abund}, with individual abundance determinations discussed in detail below.
The abundance errors are estimated by combining the dispersion around the mean with the errors introduced by the uncertainties in T$_{\mathrm{eff}}$
and $\log g$.

\subsection{The Li abundance}\label{li-abund}
\begin{figure}
\centering
\resizebox{\hsize}{!}{\includegraphics{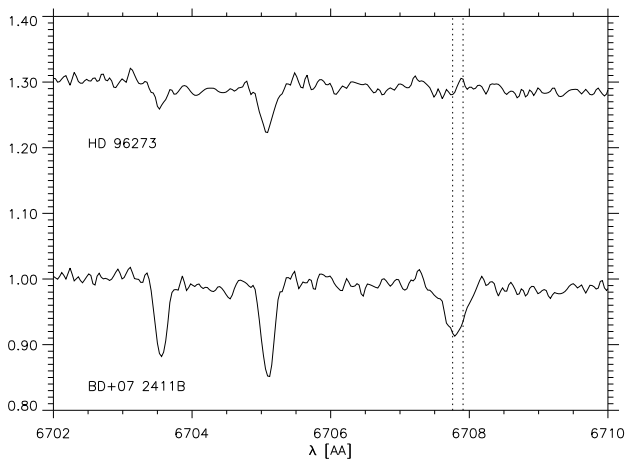}}
\caption{The region around the Li lines (positions shown by dotted lines).}
\label{fig:li}
\end{figure}

In Fig.~\ref{fig:li} we show the 
$\lambda$6708 line region for the two stars.
For \bd\ we measure EWs of the Li doublet at $\lambda\lambda$6707.761,6707.912 of 15.9~m\AA\ and 
9.6~m\AA\ respectively, giving an abundance $A(\mathrm{Li})=-10.06$. The abundances derived from the two lines differ by only 0.07dex.
 The automatic deblending performed by DAOSPEC proved inadequate, and
instead we used the IRAF\footnote{IRAF is distributed by the National Optical Astronomy Observatories,
    which are operated by the Association of Universities for Research
    in Astronomy, Inc., under cooperative agreement with the National
    Science Foundation, U.S.A.} task \emph{splot} for the deblending.
Treating the doublet as one single line yields an EW 
of 26.8~m\AA, resulting in an abundance $A(\mathrm{Li})=-10.01$. We hence adopt $A(\mathrm{Li})=-10.03$ as a good estimate of 
the lithium abundance. The dominant error source is the EW measurement, and we estimate an error of 0.10dex.
  The Li lines in the solar spectrum could not be reliably measured, so we are 
comparing with the values of \citet{grevesse+1998} to get [Li/H]~=~0.87.  

The Li abundance in \fhleo\ cannot be measured since the lines cannot be recognized, as is evident from Fig.~\ref{fig:li}. If we assume that
the $\lambda$\,6707.761 line could be as strong as 4~m\AA, we estimate an upper limit for the abundance $A(\mathrm{Li})\le-10.2$, i.e. at least
0.1 -- 0.2\,dex less
than for \bd .

Is this difference in lithium abundance significant? To address this question we note from several studies of open clusters, that the 
Li-gap \citep{boesgaard+1986} is located roughly between 6400~--~6800~K 
\citep[e.g.][for Hyades, NGC~752 and NGC~6633 respectively]{thorburn+1993,deliyannis+2000,jeffries+2002}.
However, large amounts of star-to-star scatter is generally observed at any given T$_{\mathrm{eff}}$ hotter than $\sim 6000$~K, the precise temperature
limit for the onset of the scatter depending on age. 
However, lithium depletion and the subsequent development of a Li-gap is not a straightforward 
relationship with age. In the Hyades ($700 \pm 50$~Myr) no depletion has occurred for stars cooler 
than 6200~K, while for M35 ($160 \pm 20$~Myr) depletion is already occurring down to 6000~K \citep{steinhauer+2004}.  
For the CPM star pairs investigated by \citet{martin+2002}
the situation is even more complex, although most of the systems investigated had comparable Li-abundances.  
It is believed that the rotational history of the star rather than the current rotation rate, is a determining factor
for the amount of depletion. 
On the cool side of the gap, the Li abundance usually adheres to a more straightforward
relation with age, thus the $A(\mathrm{Li})$ of \bd\ is consistent with an age somewhere between M67 \citep[4.7~Gyr;][]{pasquini+1997}
and NGC~752 \citep[1.8~Gyr;][]{deliyannis+2000}. 
If the age of the system
indeed is several Gyr, as is also suggested by Fig.~\ref{fig:tracks}, 
then it is possible that severe Li-depletion could have occurred in \fhleo\ as the star spun down to its current
low rotation rate.

\subsection{Abundances of the $\alpha$-elements}
The abundances of the electron donor elements Mg, Ca and Ti merit separate comments.  In over-active stars these elements are often 
enhanced with respect to iron, especially for low [Fe/H] stars \citep{morel+2003,dall+2005}.  
Hence these elements may act as probes of past or present magnetic activity.

The abundances of calcium was derived using 24 \ion{Ca}{i} lines. As can be seen from Table~\ref{tab:abund} the 
element is slightly overrepresented in both stars with [Ca/Fe] of 0.08 
and 0.06 for \fhleo\ and \bd\ respectively.

The magnesium abundances were found using the \ion{Mg}{i} $\lambda$\,5528.405 line, measured 
by fitting a Voigt profile to the line using \emph{splot}. While we find
[Mg/Fe]~=~0.11 for \bd\ we do not find any offset to iron for \fhleo .

The titanium abundances were measured using 24 \ion{Ti}{i} and 8 \ion{Ti}{ii} lines for \fhleo\ and  
45 \ion{Ti}{i} and 9 \ion{Ti}{ii} lines for \bd . For both stars titanium is slightly enhanced.

Given the uncertainties on the magnesium abundances, we conclude that the $\alpha$ element enhancement is probably not
larger than [$\alpha$/Fe]~=~0.06 for both stars.  Whether this enhancement could be due to past magnetic activity is doubtful,
especially given the general pattern of enhancements of other elements with respect to iron and keeping in mind the 
uncertainties on the individual abundances.

\subsection{A physically bound system}
The abundances of the two stars are similar: both are underabundant, and although \fhleo\ may be 
more underabundant than \bd, the abundances relative to iron are the same in the two stars, as evident from Table~\ref{tab:abund},
suggesting a common origin and evolution history.  Also, taking into account the uncertainties in the models, the small
abundance differences may not be real, as discussed in Sect.~\ref{abundances}.

From the spectra we also derive the radial velocities of the two components using DAOSPEC (Table~\ref{tab:params}).
The relative radial velocity between the two stars is less than 
1~\kms, i.e. a very small
velocity shift between \fhleo\ and \bd, making it even more likely that they
indeed constitute a physically bound system.

Using the Hipparcos parallax \citep{esa1997} of $8.52 \pm 1.73$\emph{mas}, we find
$d = 117$~pc, and we can reasonably adopt this as the common distance to the system.
With a separation of 8\arcsec, we find a lower limit for the physical separation of 936~AU.

\section{What caused the outburst?}\label{outburst}
Based on our analysis in Sect.~\ref{abund}, we rule out the possibility 
that an outburst occurred in a steady accretion disk of either
\fhleo\ or \bd.
While instrumental errors can never be ruled out completely, we note that in general Hipparcos has delivered
very accurate and reliable photometry.  
Also the duration of the event and the gradual fading argues against instrumental errors, since  it is unlikely that Hipparcos should have been 
giving erroneous measurements over a period of about two weeks, without all other measurements during that period being affected.
 We note that no other star in the vicinity of FH\,Leo has been flagged variable in the Hipparcos catalog \citep{esa1997}. 

We 
see at the present the following possible explanations for the Hipparcos lightcurve:
\begin{itemize}
\item Transient background or foreground  objects.
\item Magnetic interaction with unseen companion.
\item A planetary accretion event.
\item A microlensing event.
\end{itemize}

In the following we will discuss these possible scenarios and their likelihood.

\subsection{Transient background/foreground objects?}
It is in principle possible 
that the outburst is caused by another object entering  directly into the aperture, e.g. a planet, an asteroid or a supernova (SN) in a background galaxy. 
We estimated the magnitude of this object to be around $B \sim 9.9$  (i.e. even brighter than \bd) in order to result in the observed brightening.
Neither
a SIMBAD search for other objects inside the aperture, nor the FEROS guide camera images reveal any objects that could have caused such
a brightening.  Moreover, planets (and their satellites) and asteroids would produce an ``on-off'' effect as they enter and exit the aperture, not
a gradual decline, and they would move much too fast to stay within the aperture for the required time.
Deep imaging of the region around FH\,Leo may still reveal a faint distant galaxy, which could have hosted a SN bright enough to
offset the Hipparcos photometry, although we deem such a bright SN in an undiscovered galaxy very unlikely.
\begin{figure}
\resizebox{\hsize}{!}{\includegraphics{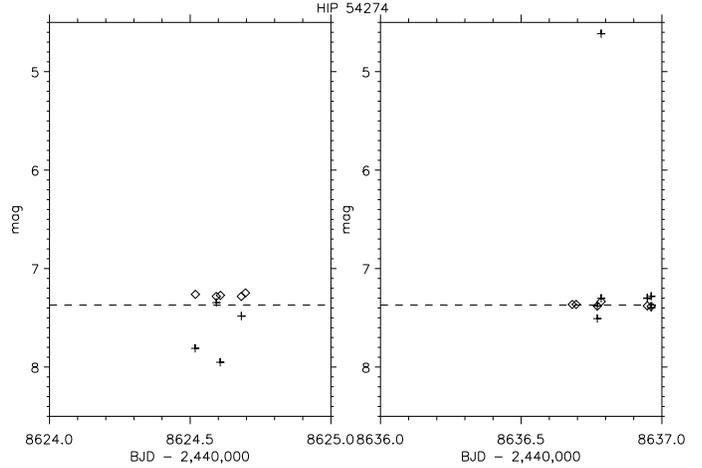}}
\caption{The Hipparcos lightcurves for HIP 54274 on the dates of the two FH\,Leo events. 
$+$: $V_T$, $\diamond$: $H$. The $B_T$ data are similar to $V_T$ and are not shown
for the sake of simplicity. Error bars are about the size of the symbols. The dashed
line show the mean magnitude in $H$.}
\label{fig:hip74}
\end{figure}
\begin{figure}
\resizebox{\hsize}{!}{\includegraphics{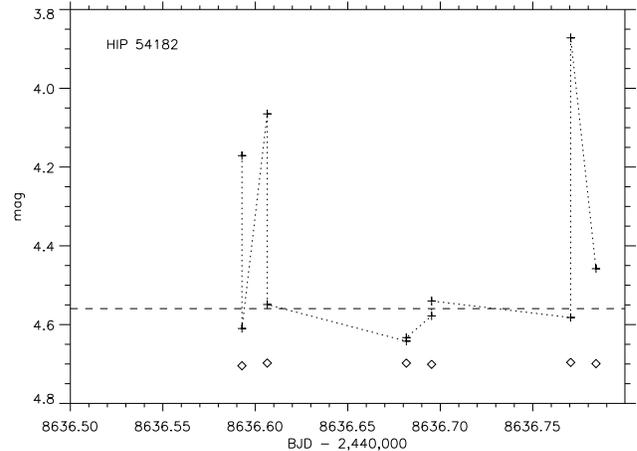}}
\caption{The Hipparcos lightcurves for HIP 54182 on the date of the second FH\,Leo event.
Symbols as Fig.~\ref{fig:hip74}. 
The dashed line show the mean magnitude in $V_T$, excluding the three largely deviating points. The $V_T$ data have been offset vertically for clarity,
and a dotted line has been drawn to better display the chronology of the data points.}
\label{fig:hip82}
\end{figure}

Another, more likely possibility is that  light from a nearby bright planet could have entered the aperture indirectly. 
During the observed brightening, Jupiter was only $\sim 6$\arcmin\ away, and it is possible that scattered light somehow
may have entered the Hipparcos instruments.   To test the likelihood of this hypothesis we examined the Hipparcos lightcurves of 
close-by stars.  
On JD\,2,448,624.6 (i.e. the time of the main event) Jupiter was at RA~=~11:05.5, DEC~=~$+$07:07, which is about 6\arcmin\
away from FH\,Leo and about the same distance from nearby \object{HIP 54274}.  The lightcurve of HIP\,54274 (Fig.~\ref{fig:hip74}) may show
a slight brightening of $\sim 0.1$\mag\ in $H$ at the time of the main event, 
and shows one spurious data point corresponding to almost 3\mag\ brightening in $V_T$ and $B_T$ at the time
of the secondary event, with Jupiter now half a degree away.  No brightening in $H$ can be seen at this epoch.

Also at the time of the second event, \object{HIP 54182} located $\sim 8$\arcmin\
from Jupiter showed a 0.7\mag\ brightening in the $V_T$ filter, while being constant in $H$ and $B_T$ (Fig.~\ref{fig:hip82}).  
This brightening is obviously of spurious nature, since the data points jump up and down with 0.5\mag\ in a few hours.
We also checked the lightcurves
of \object{HIP 54057} and \object{HIP 54331}, both of which at some point were closer than 30\arcmin\ to Jupiter, but both showed constant
lightcurves.   

Based on the fact that other Hipparcos lightcurves show similar brightenings, it is quite likely that the apparent outburst of FH\,Leo
may be caused by scattered light from Jupiter. However, the distance from Jupiter is large, and the alleged effect
seem to have a random element since not all stars in the area show similar brightenings and not necessarily at the expected time.
We note however, that caution is needed in the interpretation of Hipparcos lightcurves, since errors can be larger than what is 
suggested by the listed error bars.

A quick inspection of the distribution of Hipparcos unsolved variables on the sky, do not suggest a larger concentration near the Ecliptic plane,
as would be expected if scattered light were a major source of mistaken variability.

\subsection{Magnetic interaction with unseen companion?}\label{activity}
It is immediately obvious from the spectra that neither of the stars are over-active. The \ion{Ca}{ii} H and K lines do not show any emission cores
(Fig.~\ref{fig:caII})
and a direct comparison with the FEROS solar spectrum shows that they are both less active than the Sun.   Also the infrared Ca triplet shows no
signs of emission cores, and none of the Balmer lines show any trace of fill-in emission.  The system has also gone undetected by all the high energy
satellite missions.   

Hence, magnetic activity would need to have been confined to the past, for example induced
by some close-in companion in an eccentric orbit, bringing 
the companion into contact with the star at periastron passage, possibly spurring accretion and disruption of the 
atmosphere and the magnetic field inducing enhanced magnetic activity. The
orbital  period would have to be large enough to allow the intermittent magnetic activity
to die away completely.
As we find no traces of a companion
in the spectrum of either star, a neutron star or a giant planet are the most likely assumptions for the companion.  The observed
high scatter in the Hipparcos light curve during the outburst would then be due to flickering activity in the transient
accretion disk or due to continuous rapid flaring activity during the periastron passage. 

Chromospheric heating has been discussed by \citet{santos+2003} for \object{HD\,192263} and by \citet{shkolnik+2003}
for \object{HD\,179949}.  In these cases 
chromospheric heating occur due to
magnetic interaction between the star and a giant planet or brown dwarf in a very close, short period orbit \citep{saar+2001}. 
A similar mechanism may be at work in the FH\,Leo system, but possibly only during periastron passage.

Giant planets have been found in very eccentric orbits e.g.~\object{HD 80606} \citep{naef+2001}, which harbors a giant planet in a $e = 0.927$ orbit.
In such a case
the direct cause of the outburst could be a superflare \citep{rubenstein+2000}, where     
the snapping of field lines during a very close periastron passage could have produced enough energy for the outburst, and temporarily disrupted
any remaining field.

Another possibility is a compact companion: This scenario would be similar to an Be/X-ray transient \citep{negueruela1998}, 
which due to the lack of wind or
disk signatures, would have to be in a very eccentric long-period orbit.  However, these transients
are known to exist with early type stars, while no such objects have been found with G--type companions, which
makes this scenario rather unlikely.

With current instrumentation \citep[i.e.~HARPS;][]{harps2003} the detection of even planetary-size  companions would be possible, and
this possibility should be investigated further.

\subsection{Planetary accretion?}
Several discussions of possible planetary engulfments have been published recently \citep{siess+1999,sandquist+2002,israelian+2001,israelian+2003}.
Such an event would leave a ``polluted'' stellar atmosphere and
thus lead to enhanced abundances, especially of the volatiles \element[][6]{Li}, \element[][7]{Li}, \element[][9]{Be} and \element[][11]{B}.
Unfortunately, FEROS cannot resolve the  \element[][6]{Li}+\element[][7]{Li} blend, and it cannot reach the 
Be lines near the atmospheric UV cutoff.

The amount of energy released in a planetary engulfment process can be estimated as follows:
The outburst amplitude of 0.35\,mag corresponds to a 
ratio of energy of 1.38. Assuming solar luminosity for the quiescence star
we integrate the energy emitted during the outburst to 
$\Delta E = 10^{32}$~W. Assuming that all this energy is provided by 
the gravitational energy $E = {\rm G}\,M_{\rm star}\,m\,R_{\rm star}^{-1}$ 
of the accreted material, and taking the solar values for
the mass $M_{\rm star}$ and radius $R_{\rm star}$ of the star, we derive 
the mass $m$ of the accreted matter $m = 5\,10^{20}$~kg, which is about the
mass of a large asteroid like Pallas or Vesta.

 A scenario where \bd\ accreted a planetary-sized  companion could
explain both the Hipparcos outbursts, the possible slight overabundances found in \bd\ with respect to \fhleo, and the
presence of lithium in \bd.  One crucial test of this explanation would be to 
measure the  \element[][6]{Li}/\element[][7]{Li} isotopic ratio as done by \citet{israelian+2001,israelian+2003}, and the abundance of 
beryllium.

\subsection{Microlensing?}
The low time resolution and poorly resolved shape of the outburst as well as
the absence of color information make it impossible to judge the event as caused by 
microlensing.  However, as the system has high proper motion (-14.04, -61.85)~$\pm$~(1.81, 1.35)~mas/yr, at least the probability
for such an event exists.  Deep imaging of the region around the system might reveal a faint object
in the path of either star which was either lensed or functioned as a lens, and would thus strengthen
this interpretation of the brightening.
Note though that microlensing would not naturally explain the minor event that 
occurred 150 days later.

\section{Summary}\label{summary}

In this paper we have shown that the variable FH\,Leo cannot possibly be a CV as previously thought. We have found the
atmospheric parameters of the two stars in the system and from their derived abundances, RVs and rotation rates, shown them
to be perfectly normal main sequence stars.  

However, the outburst observed by the Hipparcos satellite remains a mystery.
While it may be caused by scattered light entering the instrument from nearby Jupiter, the effect is neither obvious  
nor reproducible when checking the lightcurves of nearby Hipparcos targets. This analysis showed however, that great care
must be exercised when interpreting Hipparcos lightcurves, since deviations much larger than the internal errors can occur.
Certainly, it would be worthwhile investigating in detail whether there exist correlations between unsolved Hipparcos variables and 
the proximity of Solar system objects to the field of view.  Our quick inspection did not suggest any correlations though.

The most severe constraints on the possible physical explanations are provided by the lack of activity on either star, which also makes it 
extremely interesting to determine the proper cause of the outburst:  If it was indeed caused by some process involving a 
very close companion inducing magnetic activity, then we have a unique possibility to gain insight into the 
time evolution of magnetic phenomena.
Hence, the explanations presented in this paper should  be assessed:  Deep imaging of the immediate surroundings
of the system could reveal a lensing object or a background galaxy.  Accurate radial velocity monitoring with high spectral resolution
could reveal any planets or compact companions in orbit around either star, and the combined spectra could provide the high signal-to-noise
needed for a measurement of the  \element[][6]{Li}/\element[][7]{Li} isotopic ratio which is a crucial test for the engulfment scenario.

\begin{acknowledgements}
This research has made use of the SIMBAD database,
operated at CDS, Strasbourg, France

Support from Funda\c{c}\~ao para a Ci\^encia e Tecnologia (Portugal)
to NCS in the form of a scholarship is gratefully acknowledged.

We are greatly indebted to the referee Dr.~D.~Bohlender for his constructive criticism.

THD wishes to thank C.~H.~F.~Melo and H.~Bruntt for helpful discussions.

\end{acknowledgements}

\bibliographystyle{bibtex/aa}
\bibliography{2813}

\begin{thebibliography}{41}
\expandafter\ifx\csname natexlab\endcsname\relax\def\natexlab#1{#1}\fi

\bibitem[{{Boesgaard} \& {Tripicco}(1986)}]{boesgaard+1986}
{Boesgaard}, A.~M. \& {Tripicco}, M.~J. 1986, \apj, 303, 724

\bibitem[{{Bruntt} {et~al.}(2004){Bruntt}, {Bikmaev}, {Catala}, {Solano},
  {Gillon}, {Magain}, {Van't Veer-Menneret}, {St{\" u}tz}, {Weiss},
  {Ballereau}, {Bouret}, {Charpinet}, {Hua}, {Katz}, {Ligni{\` e}res}, \&
  {Lueftinger}}]{bruntt+2004}
{Bruntt}, H., {Bikmaev}, I.~F., {Catala}, C., {et~al.} 2004, \aap, 425, 683

\bibitem[{{Bruntt} {et~al.}(2002){Bruntt}, {Catala}, {Garrido},
  {Rodr{\'{\i}}guez}, {St{\" u}tz}, {Knoglinger}, {Mittermayer}, {Bouret},
  {Hua}, {Ligni{\` e}res}, {Charpinet}, {Van't Veer-Menneret}, \&
  {Ballereau}}]{bruntt+2002}
{Bruntt}, H., {Catala}, C., {Garrido}, R., {et~al.} 2002, \aap, 389, 345

\bibitem[{Dall {et~al.}(2005)Dall, Bruntt, \& Strassmeier}]{dall+2005}
Dall, T.~H., Bruntt, H., \& Strassmeier, K.~G. 2005, \aap, submitted

\bibitem[{{Deliyannis} {et~al.}(2000){Deliyannis}, {Pinsonneault}, \&
  {Charbonnel}}]{deliyannis+2000}
{Deliyannis}, C.~P., {Pinsonneault}, M.~H., \& {Charbonnel}, C. 2000, in IAU
  Symposium 198, 61

\bibitem[{{Downes} {et~al.}(2001){Downes}, {Webbink}, {Shara}, {Ritter},
  {Kolb}, \& {Duerbeck}}]{downes+2001}
{Downes}, R.~A., {Webbink}, R.~F., {Shara}, M.~M., {et~al.} 2001, \pasp, 113,
  764

\bibitem[{{ESA}(1997)}]{esa1997}
{ESA}. 1997, The Hipparcos and Tycho Catalogues, ESA SP-1200

\bibitem[{{Girardi} {et~al.}(2000){Girardi}, {Bressan}, {Bertelli}, \&
  {Chiosi}}]{girardi+2000}
{Girardi}, L., {Bressan}, A., {Bertelli}, G., \& {Chiosi}, C. 2000, \aaps, 141,
  371

\bibitem[{{Gonzalez} \& {Lambert}(1996)}]{gonzalez+1996}
{Gonzalez}, G. \& {Lambert}, D.~L. 1996, \aj, 111, 424

\bibitem[{{Grevesse} \& {Sauval}(1998)}]{grevesse+1998}
{Grevesse}, N. \& {Sauval}, A.~J. 1998, Space Science Reviews, 85, 161

\bibitem[{{Halbwachs}(1986)}]{halbwachs1986}
{Halbwachs}, J.~L. 1986, \aaps, 66, 131

\bibitem[{{Israelian} {et~al.}(2001){Israelian}, {Santos}, {Mayor}, \&
  {Rebolo}}]{israelian+2001}
{Israelian}, G., {Santos}, N.~C., {Mayor}, M., \& {Rebolo}, R. 2001, \nat, 411,
  163

\bibitem[{{Israelian} {et~al.}(2003){Israelian}, {Santos}, {Mayor}, \&
  {Rebolo}}]{israelian+2003}
{Israelian}, G., {Santos}, N.~C., {Mayor}, M., \& {Rebolo}, R. 2003, \aap, 405,
  753

\bibitem[{{Jeffries} {et~al.}(2002){Jeffries}, {Totten}, {Harmer}, \&
  {Deliyannis}}]{jeffries+2002}
{Jeffries}, R.~D., {Totten}, E.~J., {Harmer}, S., \& {Deliyannis}, C.~P. 2002,
  \mnras, 336, 1109

\bibitem[{{Kazarovets} {et~al.}(1999){Kazarovets}, {Samus}, {Durlevich},
  {Frolov}, {Antipin}, {Kireeva}, \& {Pastukhova}}]{kazarovets+1999}
{Kazarovets}, A.~V., {Samus}, N.~N., {Durlevich}, O.~V., {et~al.} 1999,
  Informational Bulletin on Variable Stars, 4659, 1

\bibitem[{{Kovtyukh} {et~al.}(2003){Kovtyukh}, {Soubiran}, {Belik}, \&
  {Gorlova}}]{kovtyukh+2003}
{Kovtyukh}, V.~V., {Soubiran}, C., {Belik}, S.~I., \& {Gorlova}, N.~I. 2003,
  \aap, 411, 559

\bibitem[{{Kupka} {et~al.}(1999){Kupka}, {Piskunov}, {Ryabchikova}, {Stempels},
  \& {Weiss}}]{kupka+1999}
{Kupka}, F., {Piskunov}, N., {Ryabchikova}, T.~A., {Stempels}, H.~C., \&
  {Weiss}, W.~W. 1999, \aaps, 138, 119

\bibitem[{{Kurucz}(1993)}]{kurucz1993}
{Kurucz}, R. 1993, ATLAS9 Stellar Atmosphere Programs and 2 km/s grid.~Kurucz
  CD-ROM No.~13.~ Cambridge, Mass.: Smithsonian Astrophysical Observatory

\bibitem[{{Mart{\'{\i}}n} {et~al.}(2002){Mart{\'{\i}}n}, {Basri}, {Pavlenko},
  \& {Lyubchik}}]{martin+2002}
{Mart{\'{\i}}n}, E.~L., {Basri}, G., {Pavlenko}, Y., \& {Lyubchik}, Y. 2002,
  \apj, 579, 437

\bibitem[{{Mattei}(2004)}]{mattei2004}
{Mattei}, J.~A. 2004, Observations from the AAVSO International Database,
  private communication

\bibitem[{{Mayor} {et~al.}(2003){Mayor}, {Pepe}, {Queloz}, {Bouchy},
  {Rupprecht}, {Lo Curto}, {Avila}, {Benz}, {Bertaux}, {Bonfils}, {Dall},
  {Dekker}, {Delabre}, {Eckert}, {Fleury}, {Gilliotte}, {Gojak}, {Guzman},
  {Kohler}, {Lizon}, {Longinotti}, {Lovis}, {Megevand}, {Pasquini}, {Reyes},
  {Sivan}, {Sosnowska}, {Soto}, {Udry}, {van Kesteren}, {Weber}, \&
  {Weilenmann}}]{harps2003}
{Mayor}, M., {Pepe}, F., {Queloz}, D., {et~al.} 2003, The Messenger, 114, 20

\bibitem[{{Morel} {et~al.}(2003){Morel}, {Micela}, {Favata}, {Katz}, \&
  {Pillitteri}}]{morel+2003}
{Morel}, T., {Micela}, G., {Favata}, F., {Katz}, D., \& {Pillitteri}, I. 2003,
  \aap, 412, 495

\bibitem[{{Naef} {et~al.}(2001){Naef}, {Latham}, {Mayor}, {Mazeh}, {Beuzit},
  {Drukier}, {Perrier-Bellet}, {Queloz}, {Sivan}, {Torres}, {Udry}, \&
  {Zucker}}]{naef+2001}
{Naef}, D., {Latham}, D.~W., {Mayor}, M., {et~al.} 2001, \aap, 375, L27

\bibitem[{{Negueruela}(1998)}]{negueruela1998}
{Negueruela}, I. 1998, \aap, 338, 505

\bibitem[{{Pasquini} {et~al.}(1997){Pasquini}, {Randich}, \&
  {Pallavicini}}]{pasquini+1997}
{Pasquini}, L., {Randich}, S., \& {Pallavicini}, R. 1997, \aap, 325, 535

\bibitem[{{Piskunov}(1992)}]{piskunov1992}
{Piskunov}, N.~E. 1992, in Stellar Magnetism, ed.~Yu.~V. Glagolevsky \& I.~I.
  Romanjuk (St. Petersburg, Nauka), 92

\bibitem[{{Piskunov} {et~al.}(1995){Piskunov}, {Kupka}, {Ryabchikova}, {Weiss},
  \& {Jeffery}}]{piskunov+1995}
{Piskunov}, N.~E., {Kupka}, F., {Ryabchikova}, T.~A., {Weiss}, W.~W., \&
  {Jeffery}, C.~S. 1995, \aaps, 112, 525

\bibitem[{{Pont} \& {Eyer}(2004)}]{pont+2004}
{Pont}, F. \& {Eyer}, L. 2004, \mnras, 351, 487

\bibitem[{{Rubenstein} \& {Schaefer}(2000)}]{rubenstein+2000}
{Rubenstein}, E.~P. \& {Schaefer}, B.~E. 2000, \apj, 529, 1031

\bibitem[{{Saar} \& {Cuntz}(2001)}]{saar+2001}
{Saar}, S.~H. \& {Cuntz}, M. 2001, \mnras, 325, 55

\bibitem[{{Sandquist} {et~al.}(2002){Sandquist}, {Dokter}, {Lin}, \&
  {Mardling}}]{sandquist+2002}
{Sandquist}, E.~L., {Dokter}, J.~J., {Lin}, D.~N.~C., \& {Mardling}, R.~A.
  2002, \apj, 572, 1012

\bibitem[{{Santos} {et~al.}(2004){Santos}, {Israelian}, \&
  {Mayor}}]{santos+2004}
{Santos}, N.~C., {Israelian}, G., \& {Mayor}, M. 2004, \aap, 415, 1153

\bibitem[{{Santos} {et~al.}(2003){Santos}, {Udry}, {Mayor}, {Naef}, {Pepe},
  {Queloz}, {Burki}, {Cramer}, \& {Nicolet}}]{santos+2003}
{Santos}, N.~C., {Udry}, S., {Mayor}, M., {et~al.} 2003, \aap, 406, 373

\bibitem[{{Sbordone} {et~al.}(2004){Sbordone}, {Bonifacio}, {Castelli}, \&
  {Kurucz}}]{sbordone+2004}
{Sbordone}, L., {Bonifacio}, P., {Castelli}, F., \& {Kurucz}, R.~L. 2004,
  \memsai\ Supp., 5, 93

\bibitem[{{Shkolnik} {et~al.}(2003){Shkolnik}, {Walker}, \&
  {Bohlender}}]{shkolnik+2003}
{Shkolnik}, E., {Walker}, G.~A.~H., \& {Bohlender}, D.~A. 2003, \apj, 597, 1092

\bibitem[{{Siess} \& {Livio}(1999)}]{siess+1999}
{Siess}, L. \& {Livio}, M. 1999, \mnras, 308, 1133

\bibitem[{{Sneden}(1973)}]{sneden1973}
{Sneden}, C.~A. 1973, Ph.D.~Thesis, University of Texas

\bibitem[{{Steinhauer} \& {Deliyannis}(2004)}]{steinhauer+2004}
{Steinhauer}, A. \& {Deliyannis}, C.~P. 2004, \apjl, 614, L65

\bibitem[{Stetson \& Pancino(2005)}]{stetson+pancino2004}
Stetson, P.~B. \& Pancino, E. 2005, in preparation

\bibitem[{{Thorburn} {et~al.}(1993){Thorburn}, {Hobbs}, {Deliyannis}, \&
  {Pinsonneault}}]{thorburn+1993}
{Thorburn}, J.~A., {Hobbs}, L.~M., {Deliyannis}, C.~P., \& {Pinsonneault},
  M.~H. 1993, \apj, 415, 150

\bibitem[{{Valenti} \& {Piskunov}(1996)}]{valenti+1996}
{Valenti}, J.~A. \& {Piskunov}, N. 1996, \aaps, 118, 595

\end{thebibliography}

\end{document}